\documentclass[prd, twocolumn,12pt]{revtex4-1}

\usepackage{graphicx,amsmath,amssymb}
\usepackage{dcolumn}
\usepackage{bm}

\usepackage{hyperref}

\usepackage{graphicx,amsmath,amssymb}

\usepackage{tcolorbox}
\usepackage{amssymb}
\usepackage{listings,xcolor}
\begin{document}
\title{ A Compendium on General Relativity for Undergraduate Students}
\author{\normalsize Tushar Kanti Dey$^{1,\dag}$ and Surajit Sen${^{1,2,\S}}$\\[2ex]
$^{1}$ Centre of Advanced Studies and Innovation Lab\\ 18/27 Kali Mohan Road, Tarapur, Silchar 788003, India\\
{$^{\dag} $\tt Email: tkdey54@gmail.com}\\[2ex]
$^{2}$Department of Physics\\ Gurucharan College, Silchar 788004, India.\\
{$^{\S}$ \tt Email: ssen55@yahoo.com}\\[2ex]}

\date{\today}
\begin{abstract}
We give a pedagogical introduction of the essential features of General Theory of Relativity (GTR) in the format of an undergraduate (UG) project. A set of simple MATHEMATICA\textsuperscript{\textregistered} code is developed which enables the UG students to calculate the tensorial objects without prior knowledge of any package operation. The orbit equations of light and material particle in Minkowski and Schwarzschild spacetime are solved numerically to illustrate the crucial tests of GTR.
\end{abstract}

\maketitle

\section{Introduction}
In a reminiscence on General Theory of Relativity Einstein wrote, {\it ``I was sitting in a chair in the patent office at Bern when all of a sudden a thought occurred to me. `If a person falls freely, he will not feel his own weight'. I was startled. This simple thought made a deep impression on me. It impelled me to a theory of gravitation"}   ~\cite{Bernstein}. In contrast, Newton narrated the natural world as, ``\textit{...the whole burden of (natural) philosophy seems to consist of this - from the phenomena of motions to investigate the forces of nature, and then from these forces to demonstrate the other phenomena"} ~\cite{king}. According to him, the attraction between two gravitating objects is due to the `\emph{gravitational force}' between them, while Einstein \emph{did not} endorse it as a `force', but a manifestation of the curvature of \emph{spacetime}, an entity which is formed by soldering space and time together. Einstein's GTR is regarded as one of the foremost intellectual triumph of all time which always attracts the young undergraduate (UG) students. However, their enthusiasm is often impeded by two major obstacles: first, conceptually the tenet of GTR is completely different from Newtonian approach of gravitation, and second, the evaluation of the tensorial quantities of Riemannian geometry is very tedious and often leads to endless number of human errors. To address the first part of two major hindrances, the serious readers may go through books ~\cite{carrol2013, weinberg} and large quantity of online materials available in Internet. However, for the second part, i.e., to compute the tensorial objects, they generally consult suitable software packages to minimize the human error and save their precious time. Today there exists a plenty of free as well as commercial packages based on MATHEMATICA\textsuperscript{\textregistered} ~\cite{mathematica}, MATLAB\textsuperscript{\textregistered} ~\cite{matlab} or other software to carry out tensor algebra on computer. However, due to short span a given semester allotted for the UG project, the students of the sophomore or advanced years often face great difficulty to learn the intricacies of these package operation in particular. To address this problem, in this paper we have provided a set of MATHEMATICA\textsuperscript{\textregistered} code to calculate the tensorial objects like, Cristoffel symbol, Riemannian, Ricci, Einstein tensor and Geodesic equation which they require to learn GTR. Our approach is a easy alternative to popular MATHEMATICA\textsuperscript{\textregistered} based packages like,  MathGR\textsuperscript{\textregistered} ~\cite{mathgr}, GTRTensorII\textsuperscript{\textregistered} ~\cite{gtrtensorII} 
etc ~\cite{youtube, hartle}, which a student can workout easily without prior knowledge of package operation.\\
The remaining parts of the paper are organized as follows: In Section 2 we introduce the basic formula of Reimaniann geometry necessary to develop the subject. Section 3 gives the step by step usage of MATHEMATICA\textsuperscript{\textregistered} software to calculate the tensorial objects from some simple metric. In Section 4, we discuss the orbit equation of light and the material particle by solving the geodesics equations in Minkowski spacetime. The Einstein field equation in Schwarzschild spacetime is solved in Section 5 and the corresponding orbit equations are studied. Section 6 illustrates the numerical solution of these equations using MATHEMATICA\textsuperscript{\textregistered} code and discuss the bending of light and perihelion shift, two most coveted tests of GTR. Finally, we summarize our results and discuss outlook.

\section{Mathematical preliminaries} \label{maths}
\subsection{Key formula and equations} \label{formula}
\par
The special theory of relativity ensures frame independence of the physical laws with respect to all inertial frame of reference, while the general theory of relativity extends the same idea to include the non-inertial frame also. Any object moving under the influence of the gravitational field is essentially a accelerating object and therefore fits under that paradigm of the non-inertial frame of reference. To introduce the key effects of gravitation within framework of GTR, let us recall some basic formula of Riemannian geometry. For their detailed derivation, we refer the readers to consult any standard textbook on GTR ~\cite{carrol2013,weinberg,narlikar}.
\par
Any vector quantity $V^\mu$ (a tensor of rank one) in four dimensional space-time can be expanded as \footnote{Henceforth, the summation over the repeated indices is assumed following the Einstein convention.}
\begin{subequations}
\begin{align}
\textbf{V}(t,\textbf{r})&=\hat{e}_\mu(t,\textbf{r})V^\mu(t,\textbf{r})\label{eq1a} \\
\textbf{V}(t,\textbf{r})&=\hat{e}^\mu(t,\textbf{r})V_\mu(t,\textbf{r})\label{eq1b}
\end{align}
\end{subequations}
where $V_{\mu}$ and $V^{\mu}$ are defined as the covariant and contra-covariant vectors with corresponding basis vectors $\hat{e}_\mu$ and $\hat{e}^\mu$, respectively. We can define the two-indexed covariant and contra-variant metric tensor of rank two,
\begin{subequations}
\begin{align}
g_{\mu\nu}(t,\textbf{r})&=\hat{e}_\mu(t,\textbf{r})\hat{e}_\nu(t,\textbf{r})\label{eq2a} \\
g^{\mu\nu}(t,\textbf{r})&=\hat{e}^\mu(t,\textbf{r})\hat{e}^\nu(t,\textbf{r}),\label{eq2b}
\end{align}
\end{subequations}
which satisfies the orthogonality relation, namely,
\begin{equation}
g^{\mu\nu}g_{\mu\rho}=\delta^\nu_\rho. \label{eq3}
\end{equation}
The metric tensor have the ability to transform the contravariant vector into a covariant vector and vice versa and this operation is known as raising or lowering of indices,
\begin{subequations} \label{eq4}
\begin{align}
&V^{\mu}g_{\mu\nu}=V_\nu \qquad V_{\mu}g^{\mu\nu}=V^\nu,  \\
&V^{\mu\nu}g_{\mu\sigma}=V^\nu_{\sigma}, \quad V^{\mu}_{\nu}g_{\mu\sigma}=V_{\nu\sigma}, \nonumber \\
& V_{\mu\nu}=g_{\mu\rho}V^{\rho\sigma}g_{\sigma\nu}.
\end{align}
\end{subequations}
In general, we may have a covariant, contravariant or a mixed tensor of rank $n$, $m$ and $(n,m)$, respectively, 
\begin{equation}\label{eq5}
V_{\mu_1\mu_2\mu_3...\mu_n}, \quad V^{\mu_1\mu_2\mu_3...\mu_m}, \quad V^{\mu_1\mu_2\mu_3...\mu_m}_{\nu_1\nu_2\nu_3...\nu_n}.
\end{equation}
To develop a calculus on the spacetime continuum, it is customary to define Christoffel symbol which involves the derivative of the metric tensors with respect to spacetime coordinate $x^\mu(x^1,x^2,x^3,x^4)$,
\begin{equation}
\Gamma^\rho_{\alpha\beta}=\frac{1}{2}g^{\rho\gamma}(\frac{\partial g_{\gamma \alpha}}{\partial x^\beta}+\frac{\partial g_{\gamma \beta}}{\partial x^\alpha}-\frac{\partial g_{\alpha \beta}}{\partial x^\gamma}), \label{eq7}
\end{equation}
which is symmetric with respect to its lower indices. The covariant derivative of a second rank tensor can be defined in terms of the Cristoffel symbol, i.e.,
\begin{subequations}
\begin{align}
D_{\rho}{V_{\mu\nu}}&=\partial_\rho{V_{\mu\nu}}-\Gamma^\alpha_{\mu\rho}V_{\alpha\nu}-\Gamma^\alpha_{\nu\rho}V_{\mu\alpha}, \label{eq8a}
\\
D_{\rho}{V^{\mu\nu}}&=\partial_\rho{V^{\mu\nu}}+\Gamma^\mu_{\alpha\rho}V^{\alpha\nu}+\Gamma^\nu_{\beta\rho}V^{\mu\beta}. \label{eq8b}
\end{align}
\end{subequations}
The covariant derivative of metric tensor has vanishing value, i.e.,
\begin{equation}\label{9}
D_\alpha g_{\mu\nu}=0 \qquad  D_\alpha g^{\mu\nu}=0.
\end{equation}
The Riemann curvature tensor is defined as
\begin{equation}
(D_\mu D_\nu - D_\nu D_\mu){V_{\rho}}=R^\lambda_{\rho\nu\mu}V_\lambda, \label{eq10}
\end{equation}
where,
\begin{equation}
R^\lambda_{\mu\nu\rho}=\partial_\nu\Gamma^\lambda_{\mu\rho}-\partial_\rho\Gamma^\lambda_{\mu\nu}
+\Gamma^\eta_{\mu\rho}\Gamma^\lambda_{\eta\rho}-\Gamma^\eta_{\mu\nu}\Gamma^\lambda_{\eta\rho}, \label{eq11}
\end{equation}
and on contraction it gives the Ricci tensor and Ricci scalar, i.e.,
\begin{equation}
R_{\mu\nu}=g^{\lambda\rho}R_{\lambda\rho\mu\nu}, \qquad R=g^{\mu\nu}R_{\mu\nu},\label{eq12}
\end{equation}
respectively. It is convenient to the define a second rank tensor called `\emph{Einstein tensor}',
\begin{equation}
G_{\mu\nu}=R_{\mu\nu}-\frac{1}{2}Rg_{\mu\nu},\label{eq13}
\end{equation}
which is divergence less, namely,
\begin{equation}
D_{\mu}G^{\mu\nu}=0, \label{eq14}
\end{equation}
Thus we can write celebrated \textit{Einstein equation},
\begin{equation}
G_{\mu\nu}=-\frac{8\pi G}{c^4}T_{\mu\nu}, \label{eq15}
\end{equation}
where $T_{\mu\nu}$ be the divergence-free energy-momentum tensor of the gravitating matter which is responsible for producing the curvature in spacetime.

\subsection{Geodesic equation in curved space-time} \label{geodesic}
\par
In Newtonian mechanics, the equation of motion of a free particle ($\textbf{F}=0$) in the Eucledean space is given by,
\begin{equation}
\frac{d^2\textbf{r}}{dt^2}=0, \label{eq16}
\end{equation}
while in the special theory of relativity, which deals with the \textit{inertial frame of reference}, the dynamics is governed by the equation of motion,
\begin{equation}
\frac{d^2x^\mu}{d\tau^2}=0. \label{eq17}
\end{equation}
Here $x^\mu$ represents the spacetime coordinate in \textit{Minkowski spacetime} with $\tau$ as the body-fixed proper time.
\par
On the other hand in GTR, due to inhomogeneous character of the gravitational field, the inclusion of the acceleration becomes indispensable. Under such condition, the introduction of \textit{non-inertial frame of reference} becomes inevitable and
the equation of motion is given by so called, \textit{Geodesic Equation},
\begin{equation}
\frac{d^2x^\mu}{d\tau^2}+\Gamma^\mu_{\nu\sigma}\frac{dx^\nu}{d\tau}\frac{dx^\sigma}{d\tau}=0, \label{eq18}
\end{equation}
which completely generalizes Eq.\eqref{eq17}.
\par
Thus from the calculation point of view, the GTR involves two steps:
\begin{itemize}
\item
To solve the Einstein's equation  Eq.\eqref{eq15} to find the metric tensor $g_{\mu\nu}$ which determines the geometry of spacetime continuum.
\item
To find the solution of the geodesic equation Eq.\eqref{eq18} to know the trajectory of the point mass or massless particle in that spacetime.
\end{itemize}
In this way, GTR is essentially a metric based geometric theory of gravitation which replaces the preeminent position of \emph{force equation} advocated by Newton.

\section{MATHEMATICA\textsuperscript{\textregistered} Code for Tensorial Calculus} \label{mathematica}
\subsection{Basic flowchart} \label{flowchart}
\par
The calculation of the tensorial quantities in Riemannian geometry involves multiple derivative of the metric tensor with respect to four spacetime coordinates followed by a number of summation over the repeated indices. The purported nature of such operation often leads to endless number of errors. A suitable computer programme capable of doing symbolic calculation can perform such calculation in a error-free way within a very short span of time. In this section we present some MATHEMATICA\textsuperscript{\textregistered} notebook code to calculate the tensorial objects for any arbitrary metric tensor (Grey Box online) and list them systematically (Red Box online). The flowchart of their evaluation is given below (Download MATHEMATICA\textsuperscript{\textregistered} NOTEBOOK file \href{https://bit.ly/2rMSD6H}{here} or our \href{https://github.com/casilab/phys_ed_2020}{Github} repository):

\begin{quote}
\bf{Code I: Define list of four space-time coordinates (\verb"x1,x2,x3,x4")}:
\end{quote}
\begin{tcolorbox}
$\verb"x = List[x1,x2,x3,x4]" /.\{\verb"x1"\rightarrow{\clubsuit}, \\ \verb"x2"\rightarrow{\clubsuit}, \verb"x3"\rightarrow{\clubsuit}, \verb"x4"\rightarrow{\clubsuit}\};$
\end{tcolorbox}
where $\clubsuit$ be the unknown parameter which we need to be supplied externally.
\begin{widetext}
\begin{quote}
{\bf Code II: Define the covariant metric tensor (\verb"gcv") for a given line-element, find corresponding contra-variant (\verb"gct") metric tensor and check their orthogonality (\verb"orthg")}:
\end{quote}
\begin{tcolorbox}
\begin{flushleft}
\hspace {1.45cm}\verb"gcv = Table["$0$,\{$\alpha, 4$\}, \{$\beta,4$\}];\\
Now supply nonzero components of \verb"gcv", for example, \verb"gcv"[[$1,1$]]$=-e^{\lambda[r]}$,\, \verb"gcv"[[$2,2$]]$=-r^2$,\, \verb"gcv"[[$3,3$]]$=-r^2 \sin^2\theta$,\, \verb"gcv"[[$4,4$]]$=e^{\nu[r]}$.\\
\hspace {1.45cm}$\verb"gct = Simplify[Inverse[gcv]]; MatrixForm[gct]"$\\
\hspace {1.45cm}$\verb"orthg = FullSimplify[gcv.gct]; MatrixForm[orthg]"$\\
\end{flushleft}
\end{tcolorbox}

\begin{quote}
{\bf Code III: Calculation of Christoffel symbols ($\Gamma$) from Eq.\eqref{eq7}}:
\end{quote}
\begin{tcolorbox}
$\Gamma$\verb"a"$=$
\verb"FullSimplify[Module["\{$\alpha,\beta,\gamma,\delta$\}, \verb"Table[Sum["$\frac{1}{2}$\*\verb"gct[["$\rho,\gamma$\verb"]]"\\
\verb"(D[gcv[["$\gamma,\alpha$ \verb"]],x[["$\beta$\verb"]]+ D[gcv[["$\gamma,\beta$\verb"]], x[["$\alpha$\verb"]]-D[gcv[["$\alpha,\beta$\verb"]], x[["$\gamma$ \verb"]])]";
\end{tcolorbox}

\begin{quote}
{\bf Code IV: List of components of the Christoffel Symbol}:
\end{quote}
\begin{tcolorbox}[colback=red!5!]
\begin{flushleft}
\verb"listaffine" :=\verb" Table[If[UnsameQ["$\Gamma$\verb"[["$\alpha,\beta,\gamma$\verb"]],"$0$\verb"]," $\{$\verb"ToString[TraditionalForm["$\Gamma^{x[[\alpha]]}_{x[[\beta]],x[[\gamma]]}$ \verb"] ]", ``='', $\Gamma$\verb"a[["$\alpha,\beta,\gamma$\verb"]]" $\}$\verb"]," $\{\alpha,4\},\{\beta,4\},\{\gamma,4\}$\verb"]"\\
\verb"TableForm[DeleteCases[Flatten[listaffine,2],Null], TableSpacing" $\rightarrow \{1,1\}$\verb"]"\\
\end{flushleft}
\end{tcolorbox}

\begin{quote}
{\bf Code V: Calculation of covariant Riemann curvature tensor (\verb"Reim") from Eq.\eqref{eq11}}:
\end{quote}
\begin{tcolorbox}
\begin{flushleft}
\verb"Riem=Simplify[Module["$\{\alpha,\beta,\gamma,\delta,\rho\}$,\verb"Table[ D["$\Gamma$\verb"a[["$\rho,\alpha,\gamma$\verb"]],x[["$\beta$\verb"]]]-"
\verb"D["$\Gamma$\verb"a[["$\rho,\alpha,\beta$\verb"]],x[["$\gamma$\verb"]] ]"
\verb"+Sum[" $\Gamma$\verb"a[["$\delta,\alpha,\gamma$\verb"]]" $\Gamma$\verb"a[["$\rho,\beta,\delta$\verb"]] ," $\{\delta,4\}$ \verb"]"
\verb"-Sum[" $\Gamma$\verb"a[["$\delta,\alpha,\beta$\verb"]]" $\Gamma$\verb"a[["$\rho,\beta,\delta$\verb"]] ," $\{\delta,4\}$\verb"]", $\{\alpha,4\}$, $\{\beta,4\}$, $\{\gamma,4\}$, $\{\rho,4\}$ \verb"]]" \,  \verb"]";
\end{flushleft}
\end{tcolorbox}

\begin{quote}
{\bf Code VI: Calculation of covariant Ricci tensor (\verb"Ricicv") from Eq.\eqref{eq12}}:
\end{quote}
\begin{tcolorbox}
\begin{flushleft}
\verb"Ricicv ="\\
\verb"Simplify[Module["$\{\alpha,\beta,\gamma,\delta\}$\verb",Table[Sum[D["$\Gamma$\verb"a [["$\rho,\alpha,\beta$\verb"]],x[["$\rho$\verb"]]],"$\{\rho,4\}$\verb"]-" \verb"Sum[D["$\Gamma$\verb"a ["$\rho,\alpha,\rho$\verb"]],x[["$\beta$\verb"]]],"$\{\rho,4\}$\verb"]"
\verb"+ Sum["$\Gamma$\verb"a [["$\sigma,\alpha,\beta$\verb"]]" $\Gamma$\verb"a [["$\rho,\rho,\sigma$\verb"]]," $\{\sigma,4\}, \{\rho,4\}$\verb"]"
\verb" - Sum["$\Gamma$\verb"a [["$\sigma,\alpha,\rho$\verb"]]" $\Gamma$\verb"a [["$\rho,\beta,\sigma$\verb"]]", $\{\sigma,4\}, \{\rho,4\}$\verb"] ]  ]";
\end{flushleft}
\end{tcolorbox}

\begin{quote}
{\bf Code VII: List of components of covariant Ricci Tensor}:
\end{quote}
\begin{tcolorbox}[colback=red!5!]
\begin{flushleft}
\verb"listRicicv" :=\verb" Table[If[UnsameQ[Ricicv[["$\alpha,\beta$\verb"]]",$0$\verb"]", $\{$\verb"ToString[TraditionalForm["$R_{x[[\alpha]],x[[\beta]]}$\verb"] ]" , ``='', \verb"Ricicv[["$\alpha,\beta$\verb"]]" $\}$\verb"]", $\{\alpha,4\},\{\beta,4\}$\verb"]"\\
\verb"TableForm[DeleteCases[Flatten[listRicicv,1], Null],TableSpacing" $\rightarrow \{1,1\}$\verb"]"
\end{flushleft}
\end{tcolorbox}

\pagebreak

\begin{quote}
{\bf Code VIII: Calculation of Ricci scalar (Rc) from Eq.\eqref{eq12}}
\end{quote}
\begin{tcolorbox}
\begin{flushleft}
\verb"Rc="\\
\verb"Simplify[Module["$\{\alpha,\beta\}$,\verb"Sum[gct[["$\alpha,\beta$\verb"]]" \verb"Ricicv[["$\alpha,\beta$\verb"]]",$\{\alpha,4\},\{\beta,4\}$\verb"] ] ]";
\end{flushleft}
\end{tcolorbox}

\begin{quote}
{\bf Code IX: Calculation of Einstein covariant tensor (Gmncv) from Eq.\eqref{eq13}};
\end{quote}
\begin{tcolorbox}
\begin{flushleft}
\verb"Gmncv="\\
\verb"Simplify[Module["$\{\alpha,\beta\}$,\verb"Table[(Ricicv[["$\alpha,\beta$\verb"]]-"
$\frac{1}{2}$ \verb"gcv[["$\alpha,\beta$\verb"]] Rc)",$\{\alpha,4\},\{\beta,4\}$\verb"] ] ];"
\end{flushleft}
\end{tcolorbox}

\begin{quote}
{\bf Code X: List of components of Covariant Einstein Tensor}:
\end{quote}
\begin{tcolorbox}[colback=red!4!]
\begin{flushleft}
\verb"listGmncv" :=\verb" Table[If[UnsameQ[Gmncv[["$\alpha,\beta$\verb"]]",$0$\verb"]", $\{$\verb"ToString[TraditionalForm["$G_{x[[\alpha]],x[[\beta]]}$\verb"] ]" , ``='', \verb"Gmncv[["$\alpha,\beta$\verb"]]" $\}$\verb"]", $\{\alpha,4\},\{\beta,4\}$\verb"]"\\
\verb"TableForm[DeleteCases[Flatten[listGmncv,1], Null],TableSpacing" $\rightarrow \{1,1\}$]
\end{flushleft}
\end{tcolorbox}

\begin{quote}
{\bf Code XI: Calculation of mixed Einstein  tensor (Gmnmx)}; 
\end{quote}
\begin{tcolorbox}
\begin{flushleft}
\verb"Gmnmx="\\
\verb"Simplify[Module["$\{\alpha,\beta, \gamma\}$,\verb"Table[Sum[gct[["$\alpha, \gamma$\verb"]] Gmncv[["$\alpha,\beta$\verb"]]"$\{\gamma,4\}$,$\{\alpha,4\}, \{\beta,4\}$ \verb"] ]\, ]";
\end{flushleft}
\end{tcolorbox}

\begin{quote}
{\bf Code XII: List of components of mixed Einstein Tensor}:
\end{quote}
\begin{tcolorbox}[colback=red!4!]
\begin{flushleft}
\verb"listGmnmx" :=\verb" Table[If[UnsameQ[Gmnmx[["$\alpha,\beta$\verb"]]",$0$\verb"]", $\{$\verb"ToString[TraditionalForm["$G^{x[[\alpha]]}_{x[[\beta]]}$\verb"] ]" , ``='', \verb"Gmnmx[["$\alpha,\beta$\verb"]]" $\}$\verb"]", $\{\alpha,4\},\{\beta,4\}$\verb"]"\\
\verb"TableForm[DeleteCases[Flatten[listGmnmx,1], Null],TableSpacing" $\rightarrow \{1,1\}$]
\end{flushleft}
\end{tcolorbox}

\begin{quote}
{\bf Code XIII: Calculation of Geodesic (Orbit) equations (Geodesic) from Eq.\eqref{eq18}}:
\end{quote}
\begin{tcolorbox}
\begin{flushleft}
\verb"Geodesic="\\
\verb"Module["$\{\alpha,\beta,\gamma,s\}$\verb", Table[Simplify[D[x[["$\alpha$\verb"]]["$s$\verb"]",$\{s,2\}$\verb"]+"
\verb"Sum["$\Gamma$\verb"a[["$\alpha,\beta,\gamma$\verb"]]*D[x[["$\beta$\verb"]]["$s$\verb"]",$s$\verb"]*D[x[["$\gamma$
\verb"]]["$s$\verb"],"$s$\verb"],"
$\{\beta,4\},\{\gamma,4\}$\verb"]],"$\{\alpha,4\}$\verb"]];"
\end{flushleft}
\end{tcolorbox}

\begin{quote}
{\bf Code XIV: List of components of geodesic equations}
\end{quote}
\begin{tcolorbox}[colback=red!5!]
\begin{flushleft}
\verb"TableForm[Geodesic] // ExpandAll"
\end{flushleft}
\end{tcolorbox}
\end{widetext}
In Code XIV, each expression of the list should be equated to zero to obtain required geodesic equations of point mass and light in a given spacetime.

Using above set of codes, it is easy to evaluate various tensorial quantities for any arbitrary spacetime described by the metric tensor.

\subsection{\textbf{Application to some simple metrics}:} \label{application}
\par
The generic metric of a coordinate system is given by
\begin{equation}
ds^2=g_{\mu\nu}dx^{\mu}dx^{\nu} \label{eq19}
\end{equation}
where $g_{\mu\nu}=g_{\mu\nu}(x^1,x^2,x^3,x^4)=(\textbf{r},ct)$ be the covariant metric tensor.
\begin{enumerate}
\item
\textit{Minkowski space in cartesian coordinate}:
\par
{\bf Step I}: This spacetime is described by the line element,
\begin{equation} \label{eq20}
ds^2=dt^2-dx^2-dy^2-dz^2,
\end{equation}
where the coordinates of the system are cartesian, i.e.,
\begin{equation}
x^1 \to x, \quad x^2 \to y, \quad x^3 \to z, \quad x^4 \to t. \label{eq21}
\end{equation}
{\rm {\bf Step II}: The components of the metric tensor ($c=1$) are,}
\begin{eqnarray}
&& g_{11} \to -1, \quad g_{22} \to -1, \quad g_{33} \to -1,\nonumber \\ && g_{44} \to 1, \label{eq22}
\end{eqnarray}
which are constant. Thus in the Minkowski spacetime, the Cristoffel symbol vanishes and hence Riemann curvature, Ricci and Einstein tensors become trivial.
\item
\par
\textit{Minkowski space in spherical polar coordinate}:
\par
\textbf{Step I:} The line element in this coordinate system is given by
\begin{equation} \label{eq23}
ds^2=dt^2-dr^2-r^2 d\theta^2-r^2 \sin^2\theta d\phi^2,
\end{equation}
where coordinates are,
\begin{equation}\label{eq24}
x^1 \to r, x^2 \to \theta, x^3 \to \phi, x^4 \to t.
\end{equation}
\textbf{Step II:} The components of metric tensor directly read off from the line element Eq.\eqref{eq23} are,
\begin{eqnarray} \label{eq25}
&& g_{11} \to - 1, \quad g_{22} \to - r^2,\nonumber \\
&& g_{33} \to - r^2\sin^2\theta, \quad g_{44} \to 1.
\end{eqnarray}
\textbf{Step III:} The non-vanishing Cristoffel symbols are found to be,
\begin{eqnarray} \label{eq26}
\Gamma^\theta_{r\theta} &=& \frac{1}{r} \:, \quad \Gamma^r_{\theta\theta} = -r \:, \Gamma^r_{\phi\phi} = -r\sin^2{\theta} \: \nonumber \\
\Gamma^\theta_{\phi\phi} &=& -\cos\theta \sin\theta \:, \Gamma^\phi_{r\phi}= \frac{1}{r} \:, \nonumber \\
\Gamma^\phi_{\theta\phi} &=& \cot{\theta} \:.
\end{eqnarray}
Using the programmes (Code-V to X), the Reimann curvature, Ricci and Einstein tensors are found to be zero which indicates that in the spherical polar coordinate the Minkowski spacetime is intrinsically flat.
\item
\textit{Schwarzschild spacetime (Static, spherically symmetric and non-rotating object)}:\\
\par
\textbf{Step I:} The line element of Schwarzschild spacetime is given by
\begin{eqnarray} \label{eq27}
ds^2&=&\gamma_M d\tau^2-\frac{1}{\gamma_M} dr^2 \nonumber \\ & & -r^2(d\theta^2+\sin^2{\theta}d\phi^2),
\end{eqnarray}
where, unlike previous cases, the metric tensor is now space-dependent, i.e., $\gamma_M=\gamma_M(r)$. The coordinates of such system is same as the spherical polar coordinate system, i.e.,
\begin{equation}  \label{eq28}
x^1 \to r, \: x^2 \to \theta, \: x^3 \to \quad \phi, \: x^4 \to t.
\end{equation}
\textbf{Step II:} The components of metric tensor read off for such system are,
\begin{eqnarray}  \label{eq29}
g_{11}&=& -\frac{1}{\gamma_M}, \quad g_{22}= -r^2,\nonumber \\
g_{33}&=& -r^2\sin^2\theta, \quad g_{44}= \gamma_M,
\end{eqnarray}

\textbf{Step III:} The non-vanishing Cristoffel symbols are given by

\begin{eqnarray} \label{eq30}
\Gamma^t_{t r} &=& \frac{\gamma'_{M}}{2\gamma_{M}}, \quad \Gamma^r_{tt}=\frac{1}{2}\gamma_{M}\gamma'_{M}, \nonumber \\ \Gamma^r_{rr}&=&-\frac{\gamma'_{M}}{2\gamma_{M}}, \quad \Gamma^r_{\theta \theta}=-r \gamma_{M}, \nonumber \\ \Gamma^r_{\phi\phi}&=& -r \gamma_{M}\sin^2\theta, \quad \Gamma^\theta_{r\theta}=\frac{1}{r} \nonumber \\
\Gamma^\theta_{\phi\phi}&=&-\cos\theta \sin\theta, \quad \Gamma^\phi_{r\phi}=\frac{1}{r}, \nonumber \\
\Gamma^\phi_{\theta\phi}&=&\cot{\theta}
\end{eqnarray}

\end{enumerate}
where $\gamma_M'$ be the derivative of $\gamma_M$ with respect to $r$. The derivation of the factor $\gamma_M$ and the geodesic equation in Schwarzschild spacetime is given in Section.5. Finally we note that our code can be easily extended for spacetime for the line element containing nonzero off-diagonal terms.

\section{Geodesic equation in Minkowski spacetime} \label{geodesicmin}
\par
Before passing to GTR, let us ask `what would be the trajectory of light or a free particle in the Minkowski space-time?'. In spherical polar coordinate,
using Codes I to IV, we obtain the non-vanishing Cristoffel symbols given by Eq.\eqref{eq18}, and then,
by executing Code XIII-XIV, we get following geodesic equations,
\begin{subequations}
\begin{align}
& \frac{d^2t}{d\tau^2}=0 \label{eq31a}\\
& \frac{d^2r}{d\tau^2}-r\Bigl(\frac{d\theta}{d\tau}\Bigr)^2-r\sin^2\theta\Bigl(\frac{d\phi}{d\tau}\Bigr)^2=0 \label{eq31b}\\
&\frac{d^2\theta}{d\tau^2}+\frac{2}{r}\Bigl(\frac{dr}{d\tau}\Bigr)\Bigl(\frac{d\theta}{d\tau}\Bigr) \nonumber \\
& -\cos\theta\sin\theta\Bigl(\frac{d\phi}{d\tau}\Bigr)^2=0 \label{eq31c}\\
& \frac{d^2\phi}{d\tau^2}+\frac{2}{r}\Bigl(\frac{dr}{d\tau}\Bigr)\Bigl(\frac{d\phi}{d\tau}\Bigr) \nonumber \\
& +2\cot\theta\Bigl(\frac{d\theta}{d\tau}\Bigr)\Bigl(\frac{d\phi}{d\tau}\Bigr)=0 \label{eq31d}.
\end{align}
\end{subequations}
Eq.\eqref{eq31a} readily gives $\frac{dt}{d\tau}= constant$ and assuming it to be Lorentzian factor $\gamma$, we obtain the well known time dilation law of special theory of relativity,
\begin{equation}
\frac{dt}{d\tau}=\gamma. \label{eq32}
\end{equation}
Now if we restrict ourselves to work on a plane by setting $\theta=\pi/2$, Eq.\eqref{eq31c} becomes trivial, while Eqs.\eqref{eq31b} and \eqref{eq31d} are simplified to,
\begin{subequations}
\begin{align}
&\frac{d^2r}{d\tau^2}-r\Bigl(\frac{d\phi}{d\tau}\Bigr)^2=0,\label{eq33a} \\
&\frac{d^2\phi}{d\tau^2}+\frac{2}{r}\Bigl(\frac{dr}{d\tau}\Bigr)\Bigl(\frac{d\phi}{d\tau}\Bigr)=0,\label{eq33b}
\end{align}
\end{subequations}
respectively. From Eq.\eqref{eq33b} we obtain the angular momentum conservation law,
\begin{equation}\label{eq34}
\frac{d}{d\tau}\Bigl(r^2\frac{d\phi}{d\tau}\Bigr) =0 \quad \Rightarrow  \quad r^2\Bigl(\frac{d\phi}{d\tau}\Bigr)=h,
\end{equation}
where $h$ the conserved angular momentum. Now plucking back Eqs.\eqref{eq34} into Eqs.\eqref{eq33a} we obtain,
\begin{eqnarray}\label{eq35}
& &\frac{d}{d\tau}\Bigl(\Bigl(\frac{dr}{d\tau}\Bigr)^2+\frac{h^2}{r^2}\Bigr)=0 \nonumber \\
 & & \Rightarrow  \quad \Bigl(\frac{dr}{d\tau}\Bigr)^2+\frac{h^2}{r^2}=-\epsilon,
\end{eqnarray}
where $\epsilon$ be a constant. Finally noting the fact that,
\begin{equation}\label{eq36}
\frac{{dr}}{{d\tau }} = \frac{{dr}}{{d\varphi }}\frac{{d\varphi }}{{d\tau }} = \frac{{h}}{{r^2}}\frac{{dr}}{{d\varphi }},
\end{equation}
Eq.\eqref{eq35} gives the requisite `\textit{orbit equation}' in Minkowski spacetime in $r-\phi$ plane,
\begin{equation}
\frac{1}{{r^4}}\left({\frac{{dr}}{{d\varphi}}}\right)^2+\frac{1}{{r^2}}=-\frac{\epsilon}{{h^2}}. \label{eq37}
\end{equation}
At closest distance of approach $r=r_0$, the derivative in Eq.\eqref{eq37} vanishes and we obtain $\epsilon=-\frac{h^2}{r_0^2}$. Plucking it back into  Eq.\eqref{eq37}, we obtained its solution given by the integral,
\begin{equation}
\varphi(r)=\pm\int {\frac{{dr}}{{r^2 \sqrt{\frac{1}{{r^2}}-\frac{{1}}{{r_0^2}}}}}}. \label{eq38}
\end{equation}
The angle of deflection of the starlight due to a typical star is measured by the formula \cite{weinberg}
\begin{equation}
\hat \alpha=2\Delta\varphi-\pi, \label{eq39}
\end{equation}
where $\Delta\varphi=\varphi(r_\infty)-\varphi(r_0)$. Integration of Eq.\eqref{eq38}, in the limit $r \rightarrow \infty $, gives $\Delta\varphi=\frac{\pi}{2}$ and the angle of deflection of the starlight is found to be,
\begin{equation}
\hat \alpha=0,   \label{eq40}
\end{equation}
which indicates that the bending of light is \textit{impossible} in the Minknowski spacetime.
\section{Schwarzschild solution} \label{schwarzachild}
\subsection{Solution of Einstein's field equation} \label{fieldeqn}
In 1916, Karl Schwarzschild gave the exterior solution of the Einstein equation for a static, non-rotating and spherically symmetric object in vacuum ($T_{\mu\nu}=0$). To find the metric tensors for this spacetime, we consider the line element to be,
\begin{eqnarray}\label{eq41}
ds^2&=&B(r)dt^2-A(r)dr^2 \nonumber \\
& &-r^2(d\theta^2+\sin^2{\theta}d\phi^2).
\end{eqnarray}
Here $A(r)$ and $B(r)$ are the unknown terms, which tend to unity in the asymptotic limit ($r \rightarrow {\infty}$), are to be determined. Using Code I to X,  non-vanishing components of the Einstein tensors are given by,
\begin{subequations}\label{eq42}
\begin{align}
& r\frac{dA(r)}{dr}+A^2(r)-A(r)=0,  \\
& r\frac{dB(r)}{dr}-A(r)B(r)+B^2(r)=0.
\end{align}
\end{subequations}
The solution of Eq.\eqref{eq42} is given by,
\begin{subequations}
\begin{align}
A(r)&=\frac{1}{1-\frac{e^{C_1}}{r}}, \label{eq43a} \\
B(r)&=(1-\frac{e^{C_1}}{r})C_2, \label{eq43b}
\end{align}
\end{subequations}
where $C_1$ and $C_2$ are two constants. To get the Minkowski metric in the asymptotic limit we must choose
$C_1=ln{(2M)}, C_2=1$ and thus Schwarzschild metric reads,
\begin{eqnarray}\label{eq44}
ds^2&=&\gamma_M d\tau^2-\frac{1}{\gamma_M} dr^2 \nonumber \\
& &-r^2(d\theta^2+\sin^2{\theta}d\phi^2),
\end{eqnarray}
where the general relativistic correction factor $\gamma_M$ is given by,
\begin{equation}\label{eq45}
\gamma_M=(1-\frac{2M}{r}),
\end{equation}
which, as expected, tends to unity for $\frac{2M}{r}<<1$.

\subsection{Geodesic equation in Schwarzschild space-time} \label{geodesicsch}
\par
In this Schwarzschild spacetime, the nonvanishing Christoffel symbols are given in Eqs.\eqref{eq30}, while the corresponding geodesic equations are obtained by Codes XI and XII, respectively. In particular, unlike the geodesic equations in the Minkowski spacetime, we note that Eqs.\eqref{eq31a} and \eqref{eq31b} are generalized to
\begin{subequations}
\begin{align}
& \frac{d^2t}{d\tau^2}-\frac{\gamma'_{M}}{\gamma_{M}}\Bigl(\frac{dr}{d\tau}\Bigr)\Bigl(\frac{dt}{d\tau}\Bigr)
=0,  \label{eq46a} \\
& \frac{d^2r}{d\tau^2}-\frac{\gamma'_{M}}{2 \gamma_M} \Bigl(\frac{dt}{d\tau}\Bigr)^2
+\frac{\gamma_{M} \gamma'_M}{2}\Bigl(\frac{dr}{d\tau}\Bigr)^2 \nonumber \\
& - r\gamma_M\Bigl\{\Bigl(\frac{d\theta}{d\tau}\Bigr)^2
+\sin^2\theta\Bigl(\frac{d\phi}{d\tau}\Bigr)^2\Bigr\}=0 \label{eq46b}
\end{align}
\end{subequations}
while, Eqs.\eqref{eq31c} and \eqref{eq31d} remain unchanged. Taking $\theta=\frac{\pi}{2}$, which corresponds to the particle moving in the equatorial plane, Eq.\eqref{eq46b} is simplified to,
\begin{eqnarray}\label{eq47}
& & \frac{d^2r}{d\tau^2}- \frac{\gamma'_{M}}{2 \gamma_M} \Bigl(\frac{dr}{d\tau}\Bigr)^2
+ \frac{\gamma \gamma'_M}{2}\Bigl(\frac{dt}{d\tau}\Bigr)^2 \nonumber \\
& & - r\gamma_M\Bigl(\frac{d\phi}{d\tau}\Bigr)^2=0,
\end{eqnarray}
while Eq.\eqref{eq46a} can be written as
\begin{equation}\label{eq48}
\frac{d}{d\tau}\Bigl(ln\Bigl[\gamma_M \frac{dt}{d\tau}\Bigr]\Bigr)=0 \:
 \Rightarrow \quad \frac{dt}{d\tau}=\frac{E}{\gamma_M} ,
\end{equation}
with $E$ be a constant to be evaluated from some boundary condition. Plucking it back into Eq.\eqref{eq47} yields,
\begin{equation}
\frac{d^2r}{d\tau^2}-\frac{\gamma'_{M}}{2\gamma_{M}}\Bigl(\frac{dr}{d\tau}\Bigr)^2
+\frac{E^2 \gamma'_{M}}{2\gamma_{M}}- \frac{h^2 \gamma_{M}}{r^3}=0,  \label{eq49}
\end{equation}
which can be further simplified to
\begin{eqnarray}\label{eq50}
& & \frac{d}{d\tau}\Bigl[\frac{1}{\gamma_M}\Bigl(\frac{dr}{d\tau}\Bigr)^2-\frac{E^2}{\gamma_M} + \frac{h^2}{r^2}\Bigr]=0 \nonumber  \\
\rightarrow & & \frac{1}{\gamma_M}\Bigl(\frac{dr}{d\tau}\Bigr)^2-\frac{E^2}{\gamma_M} + \frac{h^2}{r^2}=-\epsilon.
\end{eqnarray}
Finally using Eq.\eqref{eq34} we can write Eq.\eqref{eq50} in terms of the azimuth angle $\phi$ which gives the desired `\textit{Orbit Equation}' in Schwarzschild spacetime,
\begin{equation}
\frac{1}{\gamma_M r^4 }\left( {\frac{{dr}}{{d\varphi }}} \right)^2  + \frac{1}{{r^2 }} - \frac{ E^2 }{\gamma_M h^2 } =  - \frac{\epsilon }{{h^2 }}. \label{eq51}
\end{equation}
At distance $r=r_0$, $\gamma_{M}=\gamma_{M_0}$, $(dr/d\varphi) = 0$ and Eq.\eqref{eq51} gives
\begin{equation}
E=\sqrt{\gamma_{M_0}\left(\epsilon+\frac{h^2}{r_0^2}\right)},  \label{eq52}
\end{equation}
where $\gamma_{M_0}$ be the value of $\gamma_{M}$ at $r=r_0$ i.e., $\gamma_{M_0}=1-\frac{2M}{r_0}$. Finally substituting back the value of $E$ in Eq.\eqref{eq46a}, we obtain the `time dilation' law for GTR,
\begin{equation}
\quad \frac{dt}{d\tau}=\gamma_G(r_0,M,h),  \label{eq53}
\end{equation}
where,
\begin{equation}
\gamma_G(M,r_0,h)=\frac{1}{\gamma_M}\sqrt{\gamma_{M_0}\biggl(\epsilon+\frac{h^2}{r_0^2}\biggr)}. \label{eq54}
\end{equation}
Unlike Eq.\eqref{eq32} of the Minkowski spacetime, we note that the ratio of the incremental coordinate time and proper time is function of $r, h, M_0,r_0$, respectively. The solution of Eq.\eqref{eq51} is given by,
\begin{equation}\label{eq55}
\varphi(r)=\pm \int{\frac{dr}{ r^2 (\gamma_M)^{1/2} \sqrt {\frac{E^2}{h^2 \gamma _M } - \frac{1}{r^2 } - \frac{\epsilon }{h^2 }} }},
\end{equation}
which is the Schwarzschild counterpart of Eq.\eqref{eq38} in Minkowski spacetime. In the next Section, we shall study the solution of Eq.\eqref{eq55} to find the trajectory of light and a point mass in Schwarzschild geometry, respectively.

\subsection{Effective potential in Schwarzschild geometry} \label{potentialsch}
\par
To find the effective potential in Schwarzchild geometry, from Eq.\eqref{eq50} we have \cite{carrol2013},
\begin{equation}
\frac{1}{2}E^2=\frac{1}{2}\Bigl(\frac{dr}{d\tau}\Bigr)^2+{U_{eff}^{Sch}}^2(r), \label{eq56}
\end{equation}
where the effective potential is given by
\begin{equation}
{U_{eff}^{Sch}}^2(r)=\frac{h^2}{2r^2}-\frac{h^2M}{r^3}+\frac{\epsilon}{2}\Bigl(1-\frac{2M}{r}\Bigr). \label{eq57}
\end{equation}
\begin{enumerate}
\item
\textit{Material particle ($\epsilon=1$)}: Minimizing the effective potential $U_{eff}^{Sch}(r)$ at $r=r_M$ we obtain
\begin{equation}
r_M=\frac{h^2\pm h \sqrt{h^2-12M^2}}{2M}. \label{eq58}
\end{equation}
Setting $h = 2 \sqrt{3} M$, we obtain the orbit with smallest possible radius, namely,
\begin{equation}
r_{mat}^{min}=6M. \label{eq59}
\end{equation}
for the material particle.
\item
\textit{Light ($\epsilon=0$)}: Similarly minimizing the effective potential the minimum radius for light is found to be,
\begin{equation}
r_{ph}=3M. \label{eq60}
\end{equation}
This minimal sphere of this radius is often referred as `\emph{Photon sphere}'.
\end{enumerate}
Comparing the effective potentials of the point mass ($\epsilon\neq0$) and massless particle ($\epsilon=0$) in Schwarzschild spacetime, where we note that in GTR both of them can be trapped to form a bound state.
\section{Experimental tests of GTR} \label{gtrtest}

In this section we study the trajectory of light and material particle from the orbit equation given by Eq.\eqref{eq49} and its solution Eq.\eqref{eq55}, respectively. These orbits are generally classified as the unbound ($\epsilon=0$) and bound orbit ($\epsilon\neq0$) which explain the bending of starlight and the perihelion precession of mercury, respectively.

\subsection{Motion of light in the unbound orbit - Bending of light ray:} \label{lightorbit}
To discuss the deflection of light ($\epsilon = 0$) grazing out from a star like sun at closest distance of approach ($r=r_0$), we find from Eq.\eqref{eq52},
\begin{equation}
E=\pm \frac{h \sqrt {\gamma_{M_0} }}{{r_0}}. \label{eq61}
\end{equation}
Substituting Eq.\eqref{eq61} along with $\gamma_M$ and $\gamma_{M_0}$ in Eq.\eqref{eq55} we obtain,
\begin{equation}
\Delta\varphi_{ph}=\int_{{r_0}}^\infty dr f(r; r_0, M), \label{eq62}
\end{equation}
where, $\Delta\varphi_{ph}=\varphi_{ph}(r_\infty)-\varphi_{ph}(r_0)$ and the integrand $f(r; r_0, M)$ is given by,
\begin{equation}
f(r; r_0, M)=\frac{{\sqrt{\frac{r}{{r-2M}}}}}{{r^2\sqrt{\frac{{r(r_0-2M)}}{{r_0^3(r-2M)}}-\frac{1}{{r^2}}}}}.
\label{eq63}
\end{equation}
The solution of above integral cannot be obtained exactly and there exists many approximate methods to solve it.
Here we present a direct calculation of the deflection angle using \verb"NIntegrate" program of MATHEMATICA\textsuperscript{\textregistered},

\begin{widetext}
\begin{flushleft}
\begin{tcolorbox}
\begin{equation*}
In[11]=\hat \alpha_{ph} = \left( {2\:{\rm{ NIntegrate[f, \: {{ r,\:\clubsuit}},\: \rm{ Infinity} ]\:  - \: }}\pi } \right)\frac{{180 \times\;{\rm{ }}60 \times \;{\rm{ }}60}}{\pi }{\rm{ }}
\end{equation*}
\end{tcolorbox}
\end{flushleft}
Taking the spherical body as a prototype star like sun, i.e.,
\begin{equation}\label{eq64}
M = M_{\odot} = 1.989\times 10^{30} {\rm{kg}}, \quad
r_0 = R_{\odot} = 6.95\times 10^5 {\rm{km}},  \quad
\frac{M_{\odot}G}{c^2} = 1.457 {\rm{km}},
\end{equation}
\end{widetext}
\noindent
and taking template $\clubsuit \to R_{\odot}$, the deflection angle is found to be $\hat \alpha_{ph}^{M_{\odot}}=1.75003$ arc-second. In 1919, Sir Arthur Eddington and his team verified the bending of starlight during total solar eclipse \cite{dyson}. It is worth mentioning here that, the radius of the photon sphere of the sun is $r_{ph}=3M_\odot\equiv4.425$ km obtained from Eq.\eqref{eq60} lies well within the sun, i.e., $r_{ph}<<R_{\odot}$. On the contrary, for a ultra-compact object like a black hole, the photon sphere is quite large and falls outside that object. In consequence, the bending of light for such objects is quite considerable. Finally we mention here that the bending of light leads to a unique phenomenon known as `\textit{Gravitational Lensing}', which is another spectacular outcome of GTR.

\subsection{Motion of particle in bound orbit - perihelion shift of planets}\label{particleorbit}

We finally consider the motion of a test particle ($ \epsilon \ne 0$) orbiting around the sun in an elliptical orbit. At perihelion $r=r_P$ and at aphelion $r=r_A$, $\frac{dr}{d\phi}$ vanishes in Eq.\eqref{eq51} and we get two values of the constants,
\begin{subequations}\label{eq65}
\begin{align}
\epsilon &= h^2 \frac{r_A^2 \gamma_{M_P} - r_P^2 \gamma_{M_A}}{r_P^2 r_A^2(\gamma_{M_A}-\gamma_{M_P})},   \\
E  &=\pm h\sqrt{\frac{\gamma_{M_A} \gamma_{M_P}}{\gamma_{M_A}-\gamma_{M_P}} \frac{r_A^2 -r_P^2 }{r_A^2  r_P^2 }},
\end{align}
\end{subequations}
where, $\gamma_{M_A}=1-\frac{2M}{r_A}$ and $\gamma_{M_P}=1-\frac{2M}{r_P}$, respectively. Plucking back these values in Eq.\eqref{eq55} we obtain,
\begin{equation}
\Delta\varphi_P(r)=\int_{r_P}^{r_A}f(r;M, r_P, r_A) dr \label{eq66}
\end{equation}
where $\Delta\varphi_P=\varphi(r_P)-\varphi(r_A)$ is referred as the perihelion shift of the given planet and $f$ is given by,
\begin{eqnarray}\label{eq67}
& &f(r;M, r_P, r_A) \nonumber \\
&=& \frac{\sqrt {\frac{r}{r-2 M}}}{r^2 \sqrt {\frac{(r-r_A)(r-r_P)[r r_A r_P -2 M\{r_A r_P +r(r_A + r_P)\}]}{(2 M - r) r^2 r_A^2 r_P^2}}},\nonumber \\
\end{eqnarray}
Once again the integral can be done numerically using following program of MATHEMATICA\textsuperscript{\textregistered},

\begin{widetext}
\begin{flushleft}
\begin{tcolorbox}
\begin{equation*}
\verb|In[5]|=\hat{\alpha}_P = \frac{365.25}{\clubsuit}\left( {2\:{\rm{ \verb|NIntegrate[|f,\:\{r,\:r_P}},\:{\rm{r_A\} ]\:-\:}}\pi}\right)\frac{{180 \times \;{\rm{ }}60 \times \;{\rm{}}60}}{\pi}{\rm{}}. 
\end{equation*}
\end{tcolorbox}
\end{flushleft}
\end{widetext}

\noindent
Substituting the values of $r_A, r_P$ and $\clubsuit$ (Period in days)  into above programme, we obtain the value of perihelion shift of all planets per century including Mercury. In Table-I, we have compared the perihelion shift of different planets with corresponding observational values:

\begin{widetext}
\begin{center}
{\bf Table: Perihelion shift of planets (in arc-sec per century)}
\begin{tabular}{|l|l|l|l|l|l|l|} \hline
   Planet & $r_P$ (in km) & $r_A$ (in km) & Period in days  ($\clubsuit$) & $\hat{\alpha}_{th}$& $\hat{\alpha}_{obs}$  ~\cite{carloni} \\ \hline
   Mercury & 4,60,01,200 & 6,98,16,900 & 87.97 & 42.9334 & 42.9\\
  \hline
   Venus & 10,74,76,359 & 10,89,42,109 & 224.70 & 8.6233 & 8.6 \\
  \hline
   Earth & 14,70,98,074 & 15,20,97,701 & 365.35 & 3.8364 & 3.8\\
  \hline
   Mars  & 20,66,69,000 & 24,92,09,300 & 686.97 & 1.3438 & 1.3\\
   \hline
 Jupiter & 74,05,73,600 &  81,65,20,800 & 4331.57 & .0622 & .06\\
\hline
 Saturn & 135,35,72,956 &  151,33,25,783 & 10,759.22 & .0136 & .014\\
\hline
 Uranus & 274,89,38,461 & 300,44,19,704 & 30,799.10 & .0024 & .002\\
\hline
 Neptune & 445,29,40,833 & 455,39,46,490 & 60,190.00 & .0007 & .0007\\
\hline
\end{tabular}
\end{center}
\end{widetext}

\noindent
We note that the observational value of the perihelion shift of all planets coincides with the theoretical results predicted by GTR. This striking success of GTR has established it as a complete theoretical model of gravitation.

\section{Conclusion and outlook} \label{conclusion}

This paper gives a cursory overview of General Theory of Relativity perceivable to the UG students. We presuppose that, the UG students have a rudimentary knowledge of Reimaniann geometry from their regular course and familiar with the tenets of basic MATHEMATICA\textsuperscript{\textregistered} code to undertake this study as a unsupervised (or as a project with minimal supervision) review project.
A set of easy-to-do command-line MATHEMATICA\textsuperscript{\textregistered} code is developed to solve the cumbersome tensorial quantities and to solve the orbit equations numerically. We have explicitly calculated the magnitude of the bending of light and perihelion shift of all planets including Mercury which are precisely in agreement with observational result. Apart from the UG students, our code of obtaining the orbit equation in arbitrary spacetime may be helpful for the graduate freshmen who often need to deal with wide class of metric with several nontrivial attributes.


\bibliography{gtr_education}

\begin{thebibliography}{14}
\bibitem [{\citenamefont {Wang}()}]{mathgr}%
  \BibitemOpen
  \bibfield  {author} {\bibinfo {author} {\bibfnamefont {Y.}~\bibnamefont
  {Wang}},\ }\href@noop {} {}\bibinfo {howpublished} {MathGR: A tensor and GR
  computation package to keep it simple}\BibitemShut {NoStop}%
\end{thebibliography}

\end{document}